\documentclass[aps,pra,twocolumn,showpacs]{revtex4-2}
\usepackage{amssymb}
\usepackage{color}
\usepackage{amsmath}
\usepackage{graphicx}
\usepackage{bm}
\usepackage{braket}

\begin{document}

\title{Construction of a quantum Stirling engine cycle tuned by dynamic-angle spinning}

\author{Sel\c{c}uk \c{C}akmak}
\email{selcuk.cakmak@samsun.edu.tr}
\affiliation{Department of Software Engineering, Samsun University, 55420 Samsun, Turkey}

\author{Hamid R. Rastegar Sedehi}
\email{h.rastegar@jahromu.ac.ir}
\affiliation{Department of Physics, College of Sciences, Jahrom University, Jahrom 74135-111, Iran}

\begin{abstract}
In this contribution, we investigate two coupled spins as a working substance of the quantum Stirling heat engine cycle. We propose an experimentally implementable scheme in which the cycle is driven by tuning the dipole-dipole interaction angle via dynamic-angle spinning technique in a fixed magnetic field. Realistic parameters are chosen for the proposed heat engine cycle. In addition, we aim to estimate power of the engine. To accomplish this, we focus on the microdynamics of the quantum isothermal process to predict required-time per engine cycle. The results obtained indicate that the engine produces work with high efficiency. Furthermore, the engine reaches maximum power at same point where the maximum efficiency is satisfied.
\end{abstract}

\maketitle
\section{\label{sec:intro} Introduction}
Quantum heat engines (QHEs) have recently attracted the attention of many researchers in the rapidly developing quantum mechanics. The QHE is used to produce work by using quantum matter as its working substance, which can highlight many unusual and surprising properties of quantum features~\cite{Feldman00, Scully01, Rostov03, Feldman04, Quan07, Arnaud08}. The study of the quantum analogs of classical heat engine cycles and many other generalizations have received more attention in the last decade~\cite{Scully03, Quan05, Zhang08, Dillens09, Hewgill18}.

Scientists have been proposed QHEs to extract work using various quantum systems as the working substance, such as a spin or interacting spins, two-level atom, harmonic oscillator, quantum electrodynamics systems~\cite{Quan06,Rezek06,Thomas11,Zhang14,Cakmak17,Turkpence19,Cakmak20,Lin16, Bennett20,Stefanatos17,Deffner18,Insinga20}. The role of quantum interactions between the spins have been widely investigated on the performance of the quantum heat engine cycles~\cite{Cakmak17,Turkpence19,Cakmak20}. Yin and cowokers~\cite{Yin20} have investigated an endoreversible entangled quantum Stirling engine using a 1D isotropic Heisenberg chain as working substance. They studied the influences of the exchange constant on the work output and thermal efficiency. Su et al.~\cite{Su16} have proposed a model of the quantum Otto heat engine based on the coupling via the magnetic dipole moments. The results show that the magnetic dipole-dipole interactions and the delocalized quantum states are capable of greatly enhancing the efficiency of a quantum heat engine. Also, the performance outputs of the quantum heat engine cycles under the same condition have been compared~\cite{Altintas19,Cakmak22c}.

Recently, proof-of-concept experiments of quantum Otto heat engine based on nuclear spin~\cite{Peterson19,Lisboa22}, laser coolled Cesium atom~\cite{Bouton21} and trapped ion~\cite{Zhang22} have been implemented. Also, the experimentally implementable quantum heat engine cycle driven by the dipole-dipole interaction or quadropolar interaction strengths in nuclear magnetic resonance (NMR) like setups with the low magnetic field have been proposed\cite{Cakmak22,Cakmak22b}. The results show that the loses of work and efficiency due to quantum friction for finite-time driving are negligible via tuning the dipole-dipole interaction angle in quantum adiabatic branches of the cycle.

Although the experimental implementations of the quantum Otto heat engine cycles have been studied extensively to estimation of thermalization times, quantum Stirling cycle has received much less attention. The quantum Stirling engine cycle has been studied on various quantum mechanical system~\cite{Huang02,Huang14,Chen17,Chen18,Chen20,Hamedani21,Purkait22,Papadatos22}. Additionally, the role of coupling strength on the performance of the non-regenerative quantum Stirling heat cycle under various magnitudes of the magnetic field has been recently investigated~\cite{Cakmak22c}.

In this research, we investigate dipolar coupled two spins-$1/2$ as a working substance for the quantum Stirling engine cycle. By controlling the dipole-dipole interaction angle ($\theta$), we are able to adjust the engine performance, and this could potentially be implemented in NMR-like setups experimentally. Also, we evolve the dipole-dipole interaction without making any (secular) approximation during the QSE. In addition, the engine power has approximately calculated by constructed an iterative model based on Lindblad (or Lindblad-Gorini-Kossakowski-Sudarshan) master equation~\cite{Manzano20}. Besides, the engine cycle is simulated with the realistic parameters are taken for $\mathrm{^{1}H-^{13}C}$ pair as the working substance of QSE cycle.
\section{\label{sec:theory} Theory}
\subsection{The model}
Consider a system of two spin-$1/2$ particles which are aligned in the external static magnetic field applied along the $z$-direction. The Hamiltonian of the working substance for the quantum Stirling engine includes Zeeman, J-coupling, and Dipole-dipole interactions given as follows \cite{Oliveira07,Levitt08}:
\begin{eqnarray}\label{Ham1}
 H(\theta) &=& H_Z + H_J + H_D(\theta),
\end{eqnarray}

The Zeeman splitting term of the system Hamiltonian is defined as follows:
\begin{align}\label{Zeeman}
H_Z=&-\frac{1}{2}\hbar B_0 \left( \gamma^I \sigma_{z}^I + \gamma^S \sigma_{z}^S \right)
\end{align}
where $B_0$ is magnitude of the external magnetic field that is aligned along the z-axis ($\vec{B}=B_0\vec{k}$) and $\gamma^{i}$ ($i=\left\{ I,S \right\}$) is the gyromagnetic ratio of the spin.

The $J$-coupling between the spins is given by:
\begin{align}\label{Jcoupling}
H_J=&\frac{1}{2}\pi \hbar J \left(\sigma_x^I\sigma_x^S+\sigma_y^I\sigma_y^S+\sigma_z^I\sigma_z^S \right)
\end{align}
where $J$ is the coupling constant (or interaction strength) between the $I$-spin and $S$-spin in $\mathrm{Hz}$ unit.
$J > 0$ and $ J < 0 $ correspond to the antiferromagnetic and the ferromagnetic case, respectively.

The dipole-dipole interaction between two spins, that are separated with the internuclear distance $r$, can be defined as
\begin{align}\label{DipolarCoupling}
H_D(\theta)=&-\hbar b \bigg\{ {A}_{1}+ {A}_{2}+ {A}_{3} + {A}_{4} + {A}_{5} + {A}_{6}\bigg\}.
\end{align}
The Dipolar alphabet Hamiltonian includes six-terms \cite{Goldman70,Gupta15} is given by:
\begin{subequations}\label{DipolarTerms}
\begin{align}
  {A}_{1} &=\left(1-3\cos^2{\theta}\right) \left[\frac{1}{4} \sigma_z^I \sigma_z^S \right] \\
  {A}_{2} &=\left(1-3\cos^2{\theta}\right)\left[- \frac{1}{4}\left(\sigma_{+}^I \sigma_{-}^S+\sigma_{-}^I \sigma_{+}^S \right)\right] \\
  {A}_{3} &=-\frac{3}{4}\sin{2\theta}e^{-i\phi} \left(\frac{1}{2}\sigma_{z}^I \sigma_{+}^S+\frac{1}{2}\sigma_{+}^I \sigma_{z}^S \right) \\
  {A}_{4} &=-\frac{3}{4}\sin{2\theta}e^{i\phi} \left(\frac{1}{2}\sigma_{z}^I \sigma_{-}^S+\frac{1}{2}\sigma_{-}^I \sigma_{z}^S \right) \\
  {A}_{5} &=-\frac{3}{4}\sin^2{\theta}e^{-2i\phi} \left( \sigma_{+}^I \sigma_{+}^S \right)\\
  {A}_{6} &=-\frac{3}{4}\sin^2{\theta}e^{2i\phi} \left( \sigma_{-}^I \sigma_{-}^S \right)
\end{align}
\end{subequations}
where $b=-\left(\frac{\mu_0}{4\pi}\right)\frac{\gamma^I \gamma^S \hbar}{r^3}$ dipole-dipole interaction constant. $\mu_0$ is the magnetic permeability of free space, $\theta$ represents the angle between line that connecting two spins and external magnetic field, and $\phi$ is the spherical coordinate.

\begin{figure}[!ht]\centering
\includegraphics[width=8.0cm]{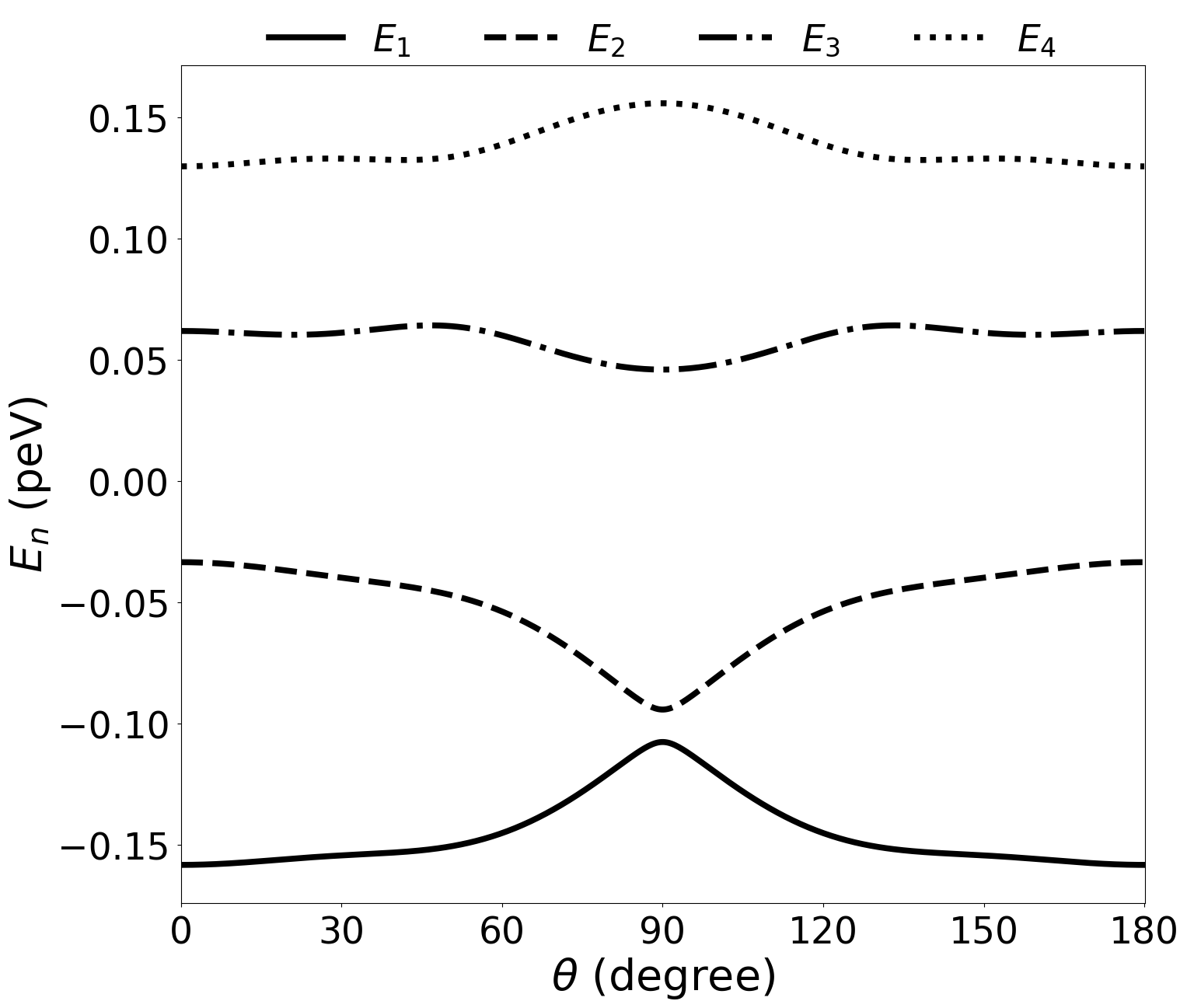}
\caption{\label{fig0} The enegy diagram of the interacted two spins with respect to the sample orientation angle ($\theta$). Because of the symmetric behavoir of the energy diagram acording the the $\theta=\pi/2$, we investigate the the cycle in range of $0$ to $\pi/2$. }
\end{figure}

\subsection{Thermalization}

Thermalization, or relaxation to thermal equilibrium state, is a phenomenon that the system reaches the thermal equilibrium with its heat bath. The history of investigation of thermalization dates back to Boltzmann and von Neumann, and many theoretical physicists have studied this problem. The problem originated in the field of non-equilibrium statistical mechanics. The time evolution of the density matrix $\rho$ can be analyzed by the so-called Lindblad-Davies maps, which generally describe the weak contact of a system with a thermal environment.
\begin{eqnarray}\label{Lindblad}
\frac{d\rho}{dt}=-\frac{i}{\hbar}\left[H,\rho\right]+ D(\rho)
 \end{eqnarray}
here $\rho$ is the density matrix of the system, $H$ is the Hamiltonian and $D(\rho)$ is a Lindblad dissipator having the Gibbs state. Also, there are many relaxation mechanisms presented in the literature~\cite{Wangsness53, Dann20, Levitt20, Manzano20}. In this research, it is considered that the perfect thermalization in Sec.~\ref{eWE} for a state which is coupled to the classical heat bath adopt the Boltzmann distribution. Hence, the density matrix of the thermal state is defined as $\rho_{T}=\exp \big\{-\hbar H/{k_B T} \big\}/Tr\left[\exp \big\{-\hbar H/{k_B T} \big\} \right]$ where $\hbar$ is the reduced Planck constant ($h/{2\pi}$) and $k_B$ is the Boltzmann constant. Additionally, in Sec.~\ref{epower}, we consider Lindblad formalism~Eq.~(\ref{Lindblad}) to calculate time-required for quantum isothermal and quantum isochoric thermodynamic processes in QSE cycle. Thus, the power of the quantum Stirling engine cycle can be estimated.

\subsection{Quantum Stirling cycle}

A quantum Stirling cycle is a quantum counterpart of classical Stirling cycle, which consists of two quantum isothermal processes and two quantum isochoric processes. The working substance of the quantum Stirling cycle is characterized quantum mechanically with the Hamiltonian given in Eq.~(\ref{Ham1}). The cycle begins from a thermal state with the temperature $T_H$ and passes through the following four stages:

Stage 1: \textit{$A \rightarrow B$} is an isothermal expansion. In this process, the system (working substance) is coupled with a heat bath at a temperature ${T}_{H}$ with the fixed low magnetic field $B$ and constant spin-spin interaction strength $J$. Also, during the all four-stages $B$ and $J$ remain unchanged. The isotermal expansion proceses implemented via tuning the $\theta$ in Eq.~(\ref{DipolarCoupling}) from the initial sample orientation $\theta_1$ to final sample orientation $\theta_2$, quasistatically. It means that the system maintains thermal equilibrium with the heat bath at any instance of time. Heat is absorbed from the bath (named ${Q}_{AB}$). The heat exchanged during the process can be calculated in terms of the change in entropy of the system, which is, in turn, associated with the change in energy of the system, and is given by
\begin{align}\label{QAB}
{Q}_{AB} =& \int_A^B T dS =T_H( S(B) - S(A) ).
\end{align}

Stage 2: \textit{$B \rightarrow C$} is a quantum isochoric process. The coupling to the heat bath is switched from a higher temperature $T_H$ to a lower temperature $T_L$. No work is done on or by the system as the other system parameters ($B$, $J$ and $D$) are fixed and heat is released (named ${Q}_{BC}$) in this process. During this stage, heat is only exchanged between the system and the bath and at the end of the stage system reaches thermal equilibrium with the cold bath at the temperature $T_L$. The amount of heat relaised contributes only to the change in internal energy of the system. This is given by
\begin{align}\label{QBC}
{Q}_{BC} =& U(C)- U(B).
\end{align}

Stage 3: \textit{$C \rightarrow D$}  is a quantum isothermal compression. The system will remain in thermal equilibrium with the cold bath at the temperature $T_L$. The isothermal compression process (stage) is implemented via tuning the $\theta$ in Eq.~(\ref{DipolarCoupling}) where  transformed back from $\theta_2$ to $\theta_1$ while the parameters $B$ and $J$ are remain unchanged during the process. The temperature of the system keeps $T_L$. Heat is released to the heat bath (${Q}_{CD}= \int_C^D T dS =  T_L ( S(D) - S(C)<0 $).

Stage 4: \textit{$D \rightarrow A$} is another quantum isochoric process. Heat is absorbed by the working substance (${Q}_{DA}=U(A)-U(D)>0$) and there is no work. In this process, the contacted heat bath temperature is swept from $T_L$ to $T_H$ and the energy levels are fixed during the isochoric transformation ($B$, $J$ and $D$ are fixed). The thermal equilibrium is satisfied at the end of the isochoric thermalization with $T_H$.

For each cycle, the extacted work from the quantum Stirling engine can be obtained considering the first law of thermodynamics. It can be defined as sum of absorbed heat and released heat $W=Q_1+Q_2$ with the efficiency $\eta=W/Q_1$. Here, $Q_1=Q_{AB}+Q_{DA}$ and $Q_2=Q_{CD}+Q_{BC}$ represents the absorbed heat and released heat, respectively. In the constructed quantum Stirling heat engine cycle above, the regenerator process is not considered. Thus, we can calculate the power of the engine precisely (see Sec.~\ref{epower}).

\section{\label{sec:results} Results}

\subsection{Work and Efficiency} \label{eWE}

We analyse the work output and efficiency characteristics of the non-regenerative quantum Stirling engine cycle which two interacting spins considered as the working substance represented quantum mechanically in Eq.~(\ref{Ham1}). The realistic parameters are taken for the $\mathrm{^{1}H-^{13}C}$ pair with the bond length is $r=1.09\AA$ (in Ethane molecule) at $T_H=100$~$\mathrm{peV/k_B}$ and $T_C=50$~$\mathrm{peV/k_B}$ for hot and cold heat baths, respectively. The gyromagnetic ratios of the nucleus are $\gamma_H/{2\pi}=42.577$ $\mathrm{MHz/T}$ and $\gamma_C/{2\pi}=10.708$ $\mathrm{MHz/T}$ (here $\mathrm{T}$ represents magnitude of the static magnetic field in Tesla unit). We set the magnitude of the external magnetic field to a fixed value of $B=1.0$~$\mathrm{mT}$. For the considered scheme, the dipolar interaction between spins has a major effect on both the work output and the efficiency of the engine. So, we don't make any (secular) aproximation on the dipolar interaction to reveal in detail its effects on the engine performance. That is the key point to obtain work with high efficiency only requires tuning the sample orientation $\theta_1 \rightarrow \theta_2$ with respect to the external magnetic field. Here, the tuning range is restricted $0$ to $\pi/2$ due to the symmetric behavoir of energy eigenvalues with respect to sample orientation, $\theta$ (see Fig.\ref{fig0}).

\begin{figure}[!ht]\centering
\includegraphics[width=8.0cm]{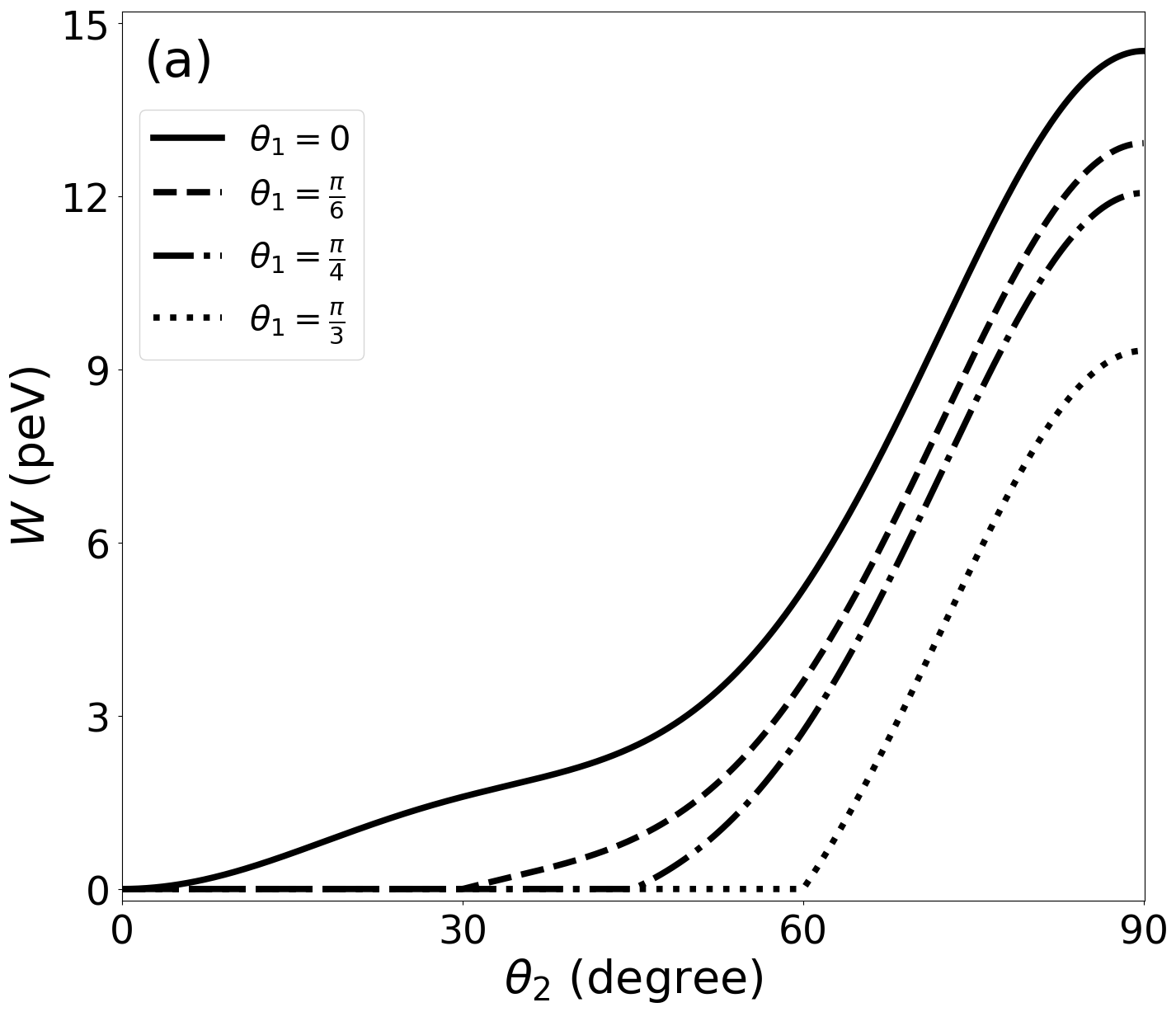}
\includegraphics[width=8.0cm]{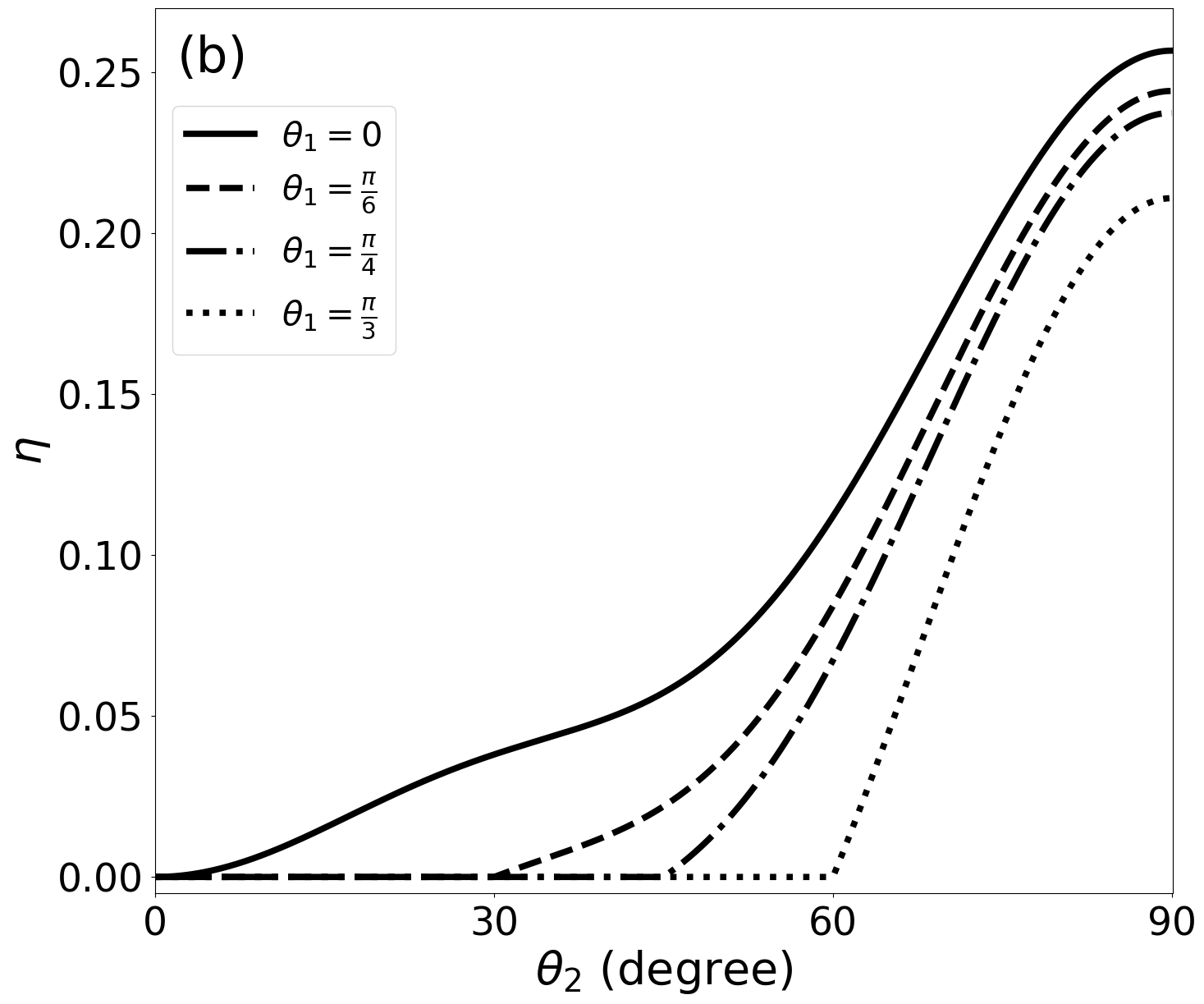}
\caption{\label{fig1} The work extracted (a) and cycle efficiency (b) of the quantum Stirling engie versus the dipole-dipole interaction angle is tuned from $\theta_1$ to $\theta_2$. In both (a) and (b), the initial sample orientation angle considered as fixed at $\theta_1=0,\frac{\pi}{6},\frac{\pi}{4},\frac{\pi}{3}$ (solid, dashed, dot-dashed, dotted) respectively and the $\theta_2$ is swept in the range of $0$ and $\pi/2$.}
\end{figure}

We first set the initial sample orientation as $\theta_1=0,\frac{\pi}{6},\frac{\pi}{4},\frac{\pi}{3}$ then the sample orientation is tuned to $\theta_2$ in the range of $0$ to $\pi/2$. The first noticeable result in Fig.~\ref{fig1}~(a) is that work extraction from the engine is possible when the $\theta_2>\theta_1$ satisfied in the considered range. The work output and cycle efficiency increase with $\theta_2$, monotonically. The maximum efficiency of the engine reaches $\eta \approx 0.25$ when $|\theta_1-\theta_2|$ is maximum (where $\theta_1=0$ and $\theta_2=\frac{\pi}{2}$ in Fig.~\ref{fig1}~(b)). Besides, the extracted work reaches the its maximum value with the maximum efficiency at same point (where $\theta_2=\frac{\pi}{2}$).

Lastly, the secular approximation is considered where $\lvert\omega_H-\omega_C\rvert>>(J-\frac{1}{2}b)$ is satisfied when the external high magnetic field applied, $B\gg 1.0$~$\mathrm{mT}$ (we set $B=1.0$~$\mathrm{T}$). Here, a high magnetic field means that, for example, the magnets in magnetic resonance imaging systems are about $1.5\mathrm{T}-3.0\mathrm{T}$. Additionally, the values can reach about $30\mathrm{T}$ in nuclear magnetic resonance setups. In this case, dipole-dipole interaction only contributes to the diagonal elements of the density matrix. Physically, this status effects the absorbed and the realized heat during the quantum isothermal and quantum isochoric stages. For the considered parameter range, the work output and the efficiency of the cycle decrease to $W\approx 0$ and $\eta\approx 0$.

\subsection{Cycle power} \label{epower}

\begin{figure}[!ht]\centering
\includegraphics[width=8.0cm]{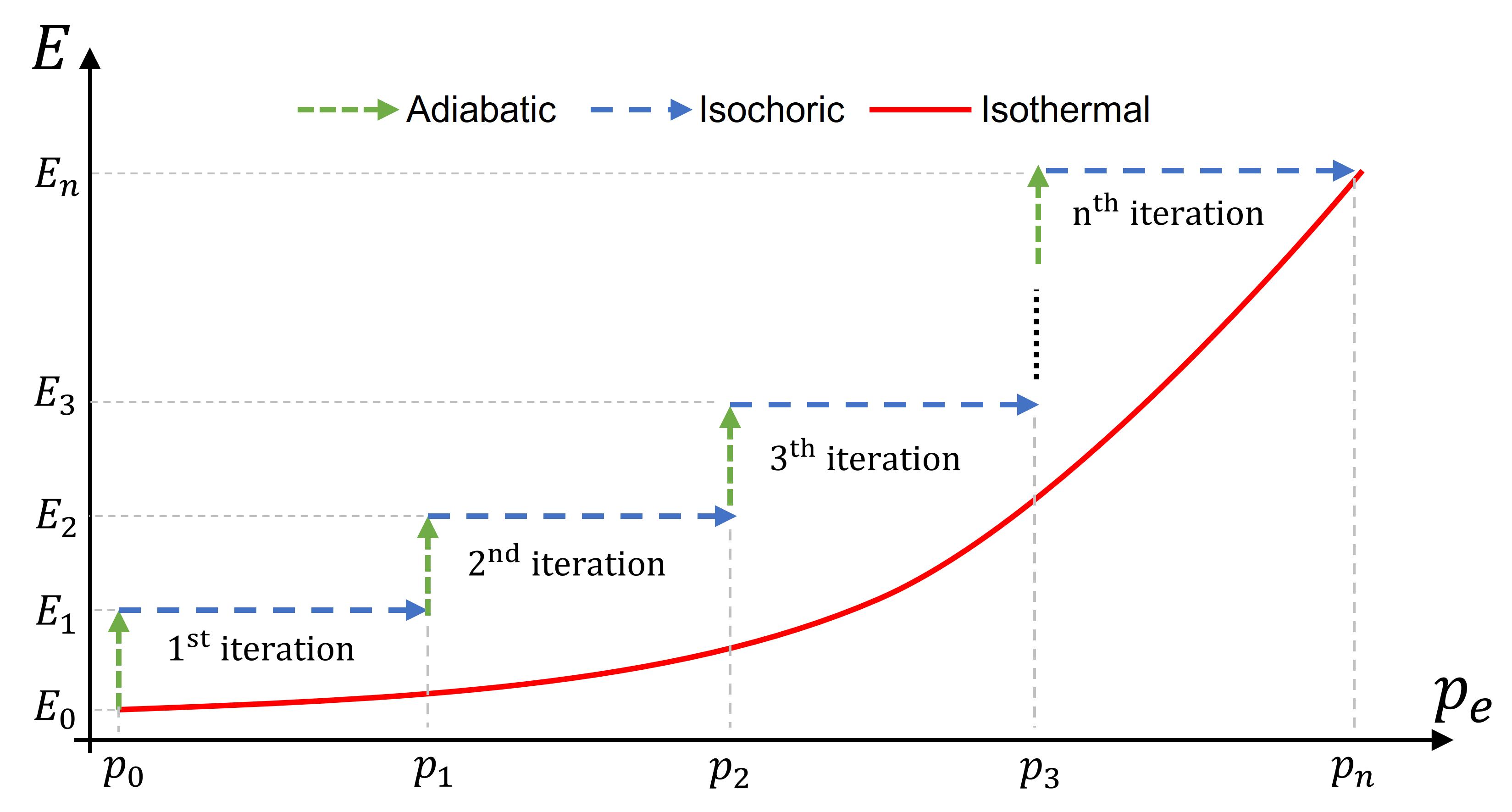}
\caption{\label{fig2} (Colour online) The excited state population versus energy for each iteration, where $E_n$ and $p_n$ are the energy and population of excited state of thermal state obtained at the end of the $n^{th}$ iteration, respectively.}
\end{figure}

In this part, we discuss the cycle power of quantum Stirling engine tuned by dipole-dipole interaction angle. First, we construct the quantum isothermal process which is considered the sum of the quantum adiabatic and quantum isochoric subprocess~\cite{Chen21} given in Fig.~\ref{fig2}. The microdynamics of the desired model analysed via the Lindblad master equation which includes all possible transitions between the energy levels of the dipolar interacted spins~\cite{Beaudoin11,Altintas13}. Also, we drive the Lindblad master equation with real system parameters to obtain the time required for a quantum isothermal process iteratively constructed.

In quantum mechanics, to calculte the closeness of two quantum states represented by the density matrices $\rho$ and $\sigma$, a common non-metric tool known as fidelity measure is used. It is defined as follows,
\begin{align}\label{fidelity}
F(\rho,\sigma)=& \left(\operatorname{Tr}\sqrt{\sqrt{\rho}\sigma\sqrt{\rho}} \right)^2.
\end{align}

\begin{figure}[!ht]\centering
\includegraphics[width=8.0cm]{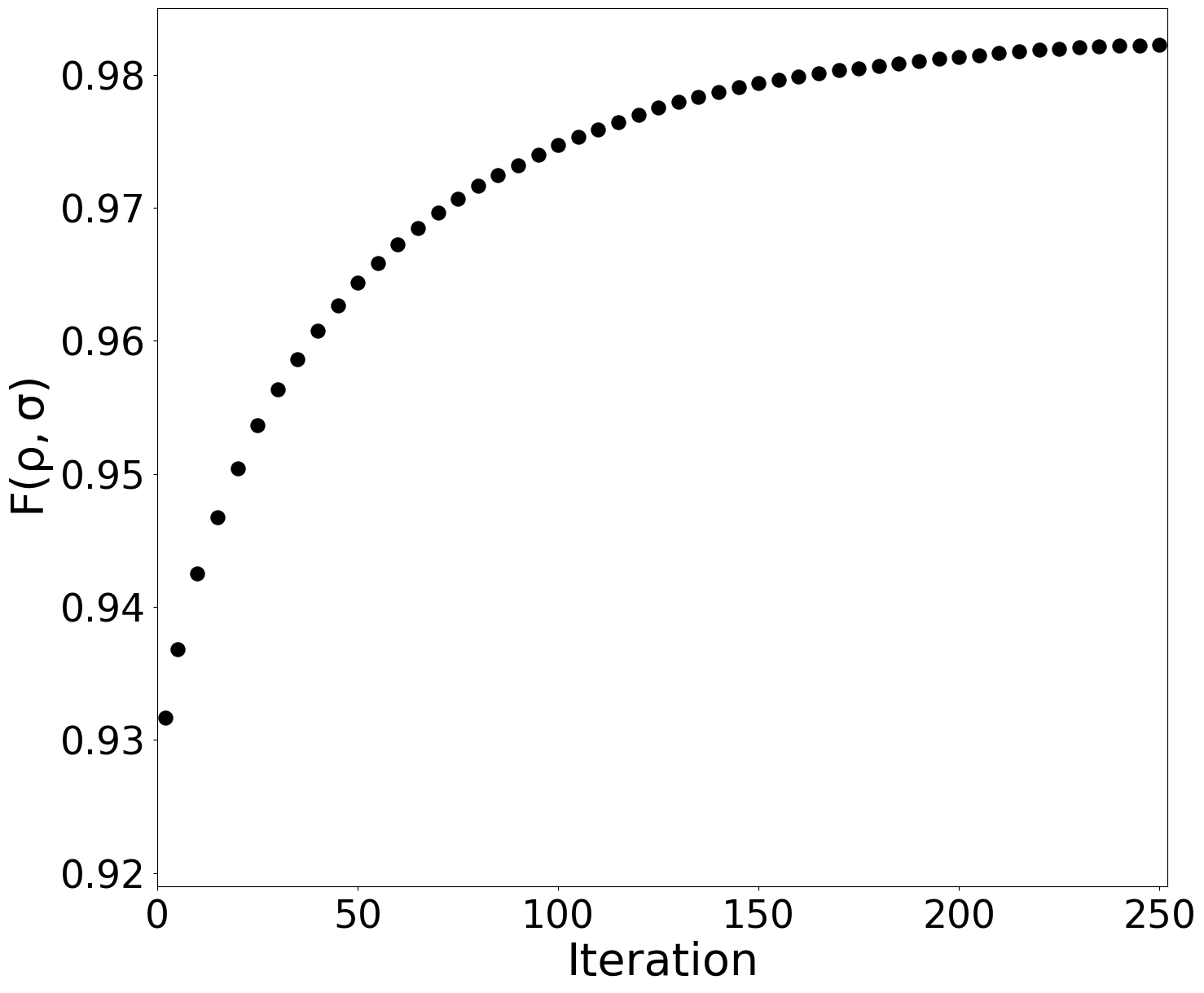}
\caption{\label{fig3} The fidelity of the state that is obtained at the end of the quantum isothermal process. Each iteration includes a quantum adiabatic and a quantum isochoric processes.}
\end{figure}

Fig.~\ref{fig3} shows the fidelity versus iteration chart where each iteration includes a quantum adiabatic and a quantum isochoric stage. In this figure, the fidelity reaches approx. $0.982$ for the constructed model. This value is comparable to the fidelity of the CNOT (Controlled NOT) gate in superconducting quantum computers which is between $0.981-0.984$~\cite{Alexander20}. So, the obtained fidelity is acceptable to estimate the time required for the quantum isothermal processes in the quantum Stirling engine. The time considered for an iteration, which includes one quantum adiabatic and one quantum isochoric process, is $\tau_{\mathrm{isothermal}}=\tau_{\mathrm{adi.}}+\tau_{\mathrm{iso.}}=100\mathrm{\mu s} + 1\mathrm{ns}\approx 100\mathrm{\mu s}$. When we focus the Fig.~\ref{fig3}, the quantum isothermal process is satisfied by the $250$ iterations. This means that the isothermal processes requires $25$ $\mathrm{ms}$, approximately.

On the other hand, the quantum isochoric process realized in a very short time ($1\mathrm{ns}$) compared to the quantum isothermal process ($25\mathrm{ms}$). Thus, the total time required per cycle is calculated as $t_{\mathrm{cycle}}\approx 50\mathrm{ms}$. The cycle power is limited by the upper value as $P_{\mathrm{cycle}}=W_{\mathrm{max}}/t_{\mathrm{cycle}}$. At the point of maximum extracted work from per cycle is ($W_{\mathrm{max}}\approx 14.51$ $\mathrm{peV}$)  (where $\theta_1=0$ and $\theta_2=\pi/2$ in Fig.~\ref{fig1}(a)) and the maximum power of the cycle reaches $P_{\mathrm{cycle}}\approx 4.65 \times 10^{-29}$~$\mathrm{J.s^{-1}}$. It should be noted that this value is only for one $\mathrm{^{1}H-^{13}C}$ pair.

\section{\label{sec:results} Conclusions}
In the presented study, we constructed the quantum Stirling engine by tuning the dipole-dipole interaction angle between two coupled spins. The sample orientation with respect to the external static magnetic field was changed from an initial angle ($\theta_1$) to a desired angle ($\theta_2$) in order to extract work from the engine. The obtained results show that it is possible to drive a Stirling cycle with positive work output and high efficiency (near Carnot efficiency, $\eta_C=1-T_C/T_H=0.5$) by only tuning the sample orientation at a fixed low magnetic field. We fully observed the dipolar interaction effects on the cycle performance without applying any secular approximation.

The work and efficiency increase when the sample orientation is tuned to a wider range, $|\theta_1-\theta_2|=\pi/2$. Additionally, we constructed a model to estimate the required time per cycle of the quantum Stirling engine. With this model, we can calculate the cycle power ($P_{\mathrm{cycle}}$) when the engine reaches its maximum work output. Furthermore, we have investigated the engine behavior under high magnetic field approximation (secular approximation), $B=1.0$~$\mathrm{T}$, where the engine is useless ($W\approx 0$) in the realistic parameter range considered.

Additionally, most conventional NMR setups have a fixed high magnetic field in the $z$-direction, and there is no way to tune it for a specific value. Furthermore, it is difficult to generate and control the high magnetic field applied to the working substance (sample) using recent scientific tools. In contrast, we propose tuning only the sample orientation to drive the quantum Stirling cycle and extract work from the cycle. An advanced experimental technique known as dynamic-angle spinning (DAS) allows for the tuning of the sample orientation in NMR setups~\cite{Mueller11}. Therefore, the experimental demonstration of the QSE can be realized using upgraded NMR-like setups.

\end{document}